\documentclass[Royal, doublespace, sageh, times]{sagej}

\usepackage{moreverb,url, csquotes, amsmath, textcomp, braket}

\usepackage[colorlinks,bookmarksopen,bookmarksnumbered,citecolor=red,urlcolor=red]{hyperref}

\hypersetup{
    colorlinks=true,
    linkcolor=blue,
    citecolor = blue,
    filecolor=black,      
    urlcolor=blue,
}

\newcommand\BibTeX{{\rmfamily B\kern-.05em \textsc{i\kern-.025em b}\kern-.08em
T\kern-.1667em\lower.7ex\hbox{E}\kern-.125emX}}

  {\begin{list}{}%
          {\setlength{\leftmargin}{#1}}%
          \item[]%
  }
  {\end{list}}

\def\volumeyear{2019}

\begin{document}

\runninghead{Alexandros Haridis}

\title{The Topology of Shapes Made with Points}

\author{Alexandros Haridis\affilnum{1}}

\affiliation{\affilnum{1}Department of Architecture, Massachusetts Institute of Technology, USA}

\corrauth{Alexandros Haridis, 77 Massachusetts Avenue, Room 10-303, Cambridge, MA 02139, USA.}

\email{charidis@mit.edu}

\begin{abstract}
In architecture, city planning, visual arts, and other design areas, shapes are often made with points, or with structural representations based on point-sets. Shapes made with points can be understood more generally as finite arrangements formed with elements (\emph{i.e.} points) of the algebra of shapes $U_i$, for $i = 0$. This paper examines the kind of topology that is applicable to such shapes. From a mathematical standpoint, any \enquote{shape made with points} is equivalent to a finite space, so that topology on a shape made with points is no different than topology on a finite space: the study of topological structure naturally coincides with the study of preorder relations on the points of the shape. After establishing this fact, some connections between the topology of shapes made with points and the topology of \enquote{point-free} pictorial shapes (when $i > 0$) are discussed and the main differences between the two are summarized.
\end{abstract}

\keywords{Shape with Points, Finite Order Topology, T$_0$-space, Structural Description, Mathematics of Shapes}

\maketitle

\newpage

\section{Mathematics of shapes}

\noindent In a previous investigation (\cite{R3}), topology was studied on (point-free) shapes made with basic elements of the algebras of shapes $U_i$, when $i > 0$. What was left undiscussed is the topology applicable to shapes made with points, \emph{i.e.} when $i = 0$. That is the subject of this present paper.

\section{Shapes made with points}
Before expanding on the main topic of this paper, it is useful to give a clear idea of what is meant by a \emph{point} and a \emph{shape made with points}. 

A well-known approach is to consider the point as an infinitesimally small entity or an abstract (immaterial) member of a set\textemdash this is common, for example, in mathematics. More generally though, a point can be considered as a representation or surrogate of a concrete (material) object---an object that has physical presentation, can be seen or touched. We see this many times, for example, in physics when modeling the motion of physical objects and also during composition with shapes in drawings and models in architecture and the visual arts. 

What are the properties of a \enquote{concrete/material object} when we say that it behaves like a point? First of, the notion of point cannot be associated with a specific mark or material figure. A point is agnostic of how it looks. This is nicely expressed in the following quote by the Russian painter W. Kandinsky:

\begin{displayquote}
\enquote{Externally, the point may be defined as the smallest elementary form, but this definition is not exact. It is difficult to fix the exact limits of \lq smallest form.\rq\; The point can grow and cover the entire ground plane unnoticed, then, where would the boundary between point and plane be?... In its material form, the point can assume an unlimited number of shapes \cite[pp.~29-31]{R6}.}
\end{displayquote}

\begin{figure}[h]
\centering
\includegraphics[width=\textwidth]{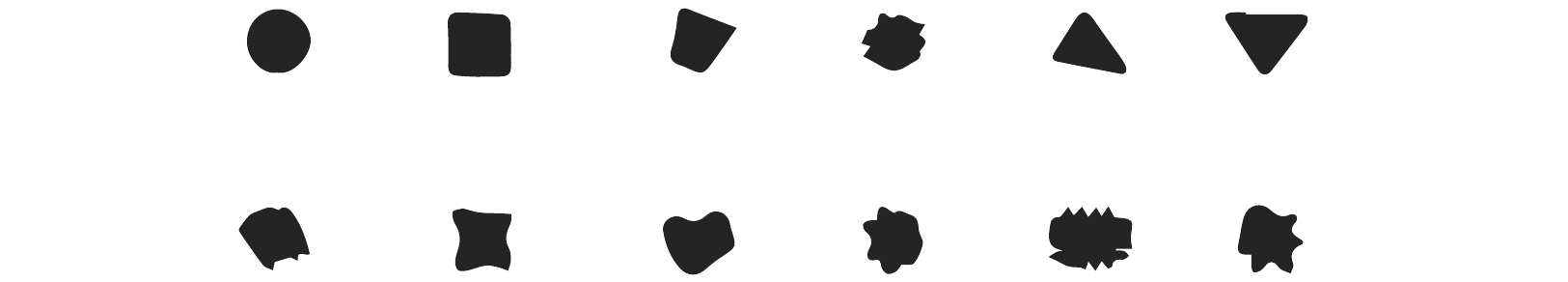}
\caption{Redrawn from \cite[p.~31]{R6}.}
\end{figure}

\noindent Any one of the shapes in Figure 1, for example, may equally function as a point. And it is only by convention that a point is often represented as a small circle (void or solid). Thus, an object that behaves as a point must have certain properties that have nothing to do with the way it looks. 

An object is said to behave like a point whenever certain assumptions have been made about how one interacts with the object and how the object is supposed to interact with other objects, when put together in a spatial composition. These assumptions (to be explained shortly) are formally captured in the mathematical framework of the algebra of points $U_0$, which is part of the shape algebras $U_i$ ($i \geq 0$) invented for purposes of calculation with \emph{shape grammars}. 

Points are the basic elements in an algebra $U_0$\textemdash lines, planes, and solids, are the rest of the elements in the algebras when $i > 0$. The algebraic properties of points, and their relation to the other basic elements in the series, are described in detail elsewhere (e.g. \cite{R12} provides a comprehensive coverage). Intuitively, the basic properties of points can be summarized as follows. A point is an element without proper nonempty parts. Unlike with lines, planes or solids, one cannot perceptually recognize parts in a point; a point has one part only, namely, itself. Symbols or indivisible units, such as \enquote{0}s and \enquote{1}s, or \enquote{a}s, \enquote{b}s and \enquote{c}s, used in linguistics and logic, behave similarly. Just like symbols in a set, points stay mutually impenetrable when put together in a spatial composition. They are meant to stay distinct and exclude one another, unless if they are identical, in which case (and only in this case) they fuse into one.

\begin{figure}[b]
\includegraphics[scale=0.33]{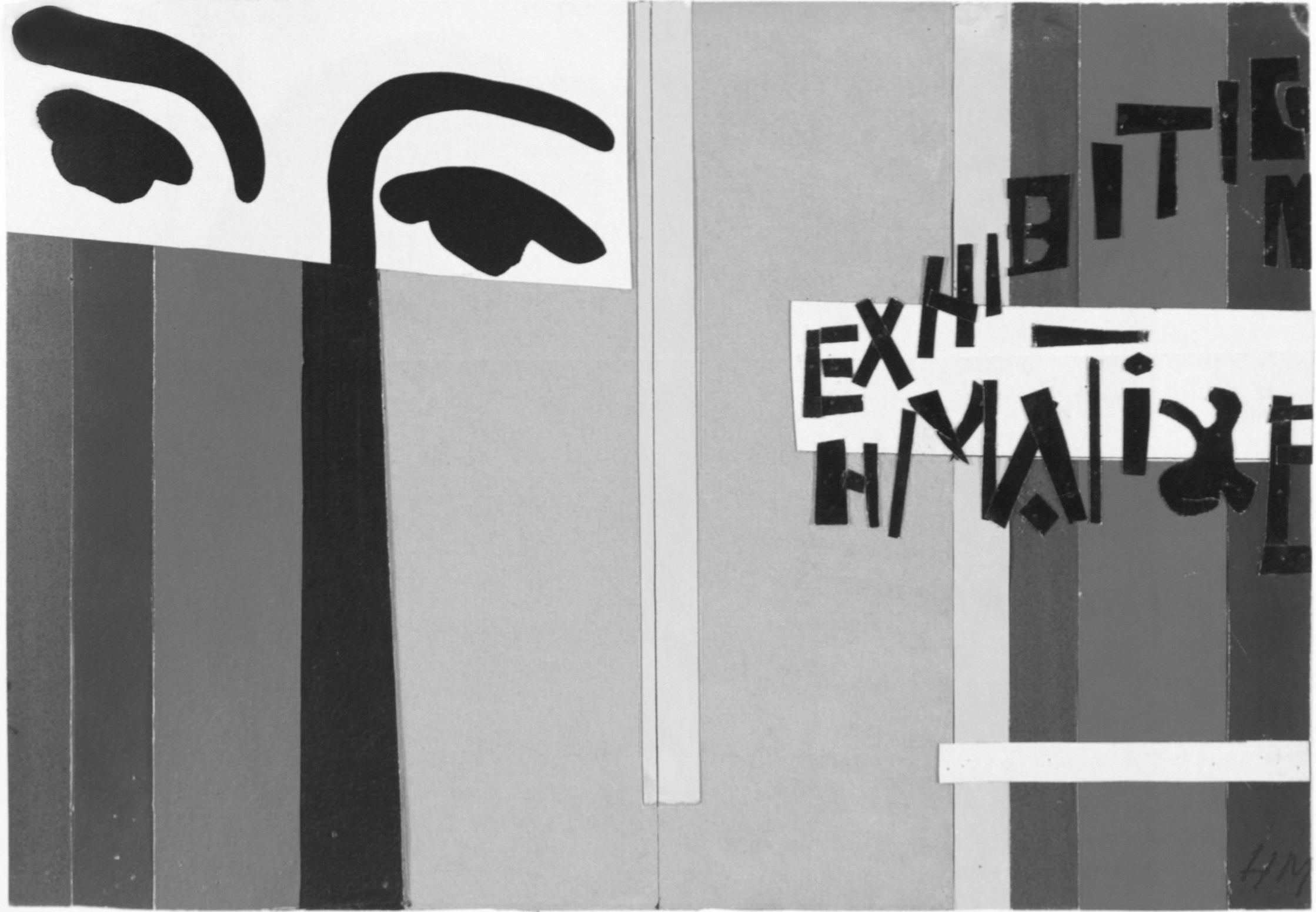}
\caption{Design for cover of exhibition catalogue \enquote{H. Matisse,} by Henri Matisse, 1951, gouache on paper, cut-and-pasted, 27.0 x 40.0 cm (\textcopyright $\;$ 2019 Succession H. Matisse / Artists Rights Society (ARS), New York).}
\end{figure}

By a \enquote{shape made with points}, we understand any finite arrangement formed with spatial elements of any form, size or shape, that function as points. One spatial example coming from the visual arts is the concept of the \emph{collage}. A collage is a type of art where one sticks different elements together to form a composition, for example, the one in Figure 2, made with paper cut-outs. The elements in a collage stay distinct from one another and do not materially fuse when combined. The elements may fuse in observation, when one looks at the resulting arrangement, and give rise to emerging forms and meanings; one may say, for example, that in Figure 2 there is a shape that looks like an \enquote{eyebrow} which extends downwards to become the \enquote{nose} of a partially visible face. Nonetheless, as far as the interaction is concerned, if one tries to touch the collage, to rearrange its elements in some other way, only the independent compositional elements can be manipulated.

\begin{figure}[b]
\includegraphics[width=\textwidth]{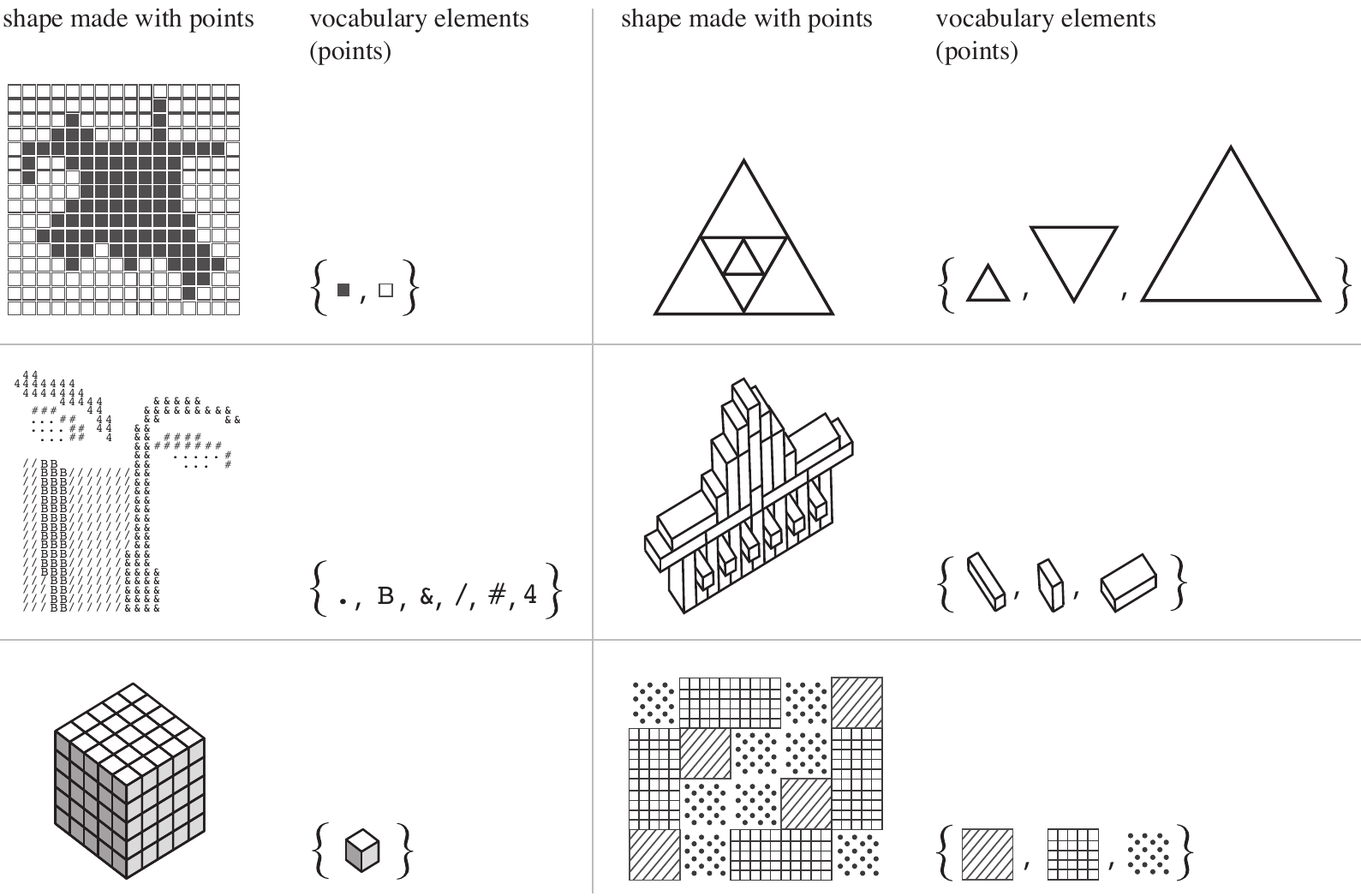}
\caption{Examples of shapes made with points. Each shape comes with the vocabulary elements (points) out of which it is composed of.}
\end{figure}

Besides collages, one can think of many other kinds of shapes, used in a wider range of areas, formed with elements that are made to have point-like behavior. While the material appearance of the elements is different each time depending on the application and use, their point-like properties are uniformly the same. Some examples are: building models, physical or digital, for architectural design determined by a \enquote{kit of parts;} Froebel-like, or similarly, Lego type designs; digital images, ASCII computer artworks and voxel descriptions of three-dimensional form; hatching systems for filling discrete regions in the plane; spatial configurations made with primitive components and relations; electrical circuit diagrams, and other graph-based descriptions of form, and so on. Figure 3 shows some such examples of shapes made with points.

Recall that when shapes are treated as \enquote{point-free}, i.e. unanalyzed, objects (when made with basic elements of algebras $U_i$ when $i > 0$), they can be interpreted into parts in indefinitely many ways, independently of what pieces were originally used to make them (\cite{R12}). In contrast, shapes made with points cannot be interpreted into parts in ways other than what their points allow. By a \emph{part} of a shape made with points, one essentially means a \emph{subset of points} of the shape. The possible parts that can be recognized and manipulated are only those that can be formed by combining the points that were originally used to form the shape: parts are subsets of points, and no more. This is explained in a more graphical manner in Figure 4. The figure shows a shape made with points (in this case a motif, extracted from the collage in Figure 2) along with a set of visible and possible parts and a set of visible but non possible parts. While the non possible parts are embedded in the shape\textemdash we can see them\textemdash, it is as if they do not exist from an interpretative standpoint: we cannot combine points to form these parts. (Obviously, one can cut the elements themselves to get a new revised set of points, but the intended argument here with respect to perceptual interpretation remains the same.)

\begin{figure}[b]
\includegraphics[width=\textwidth]{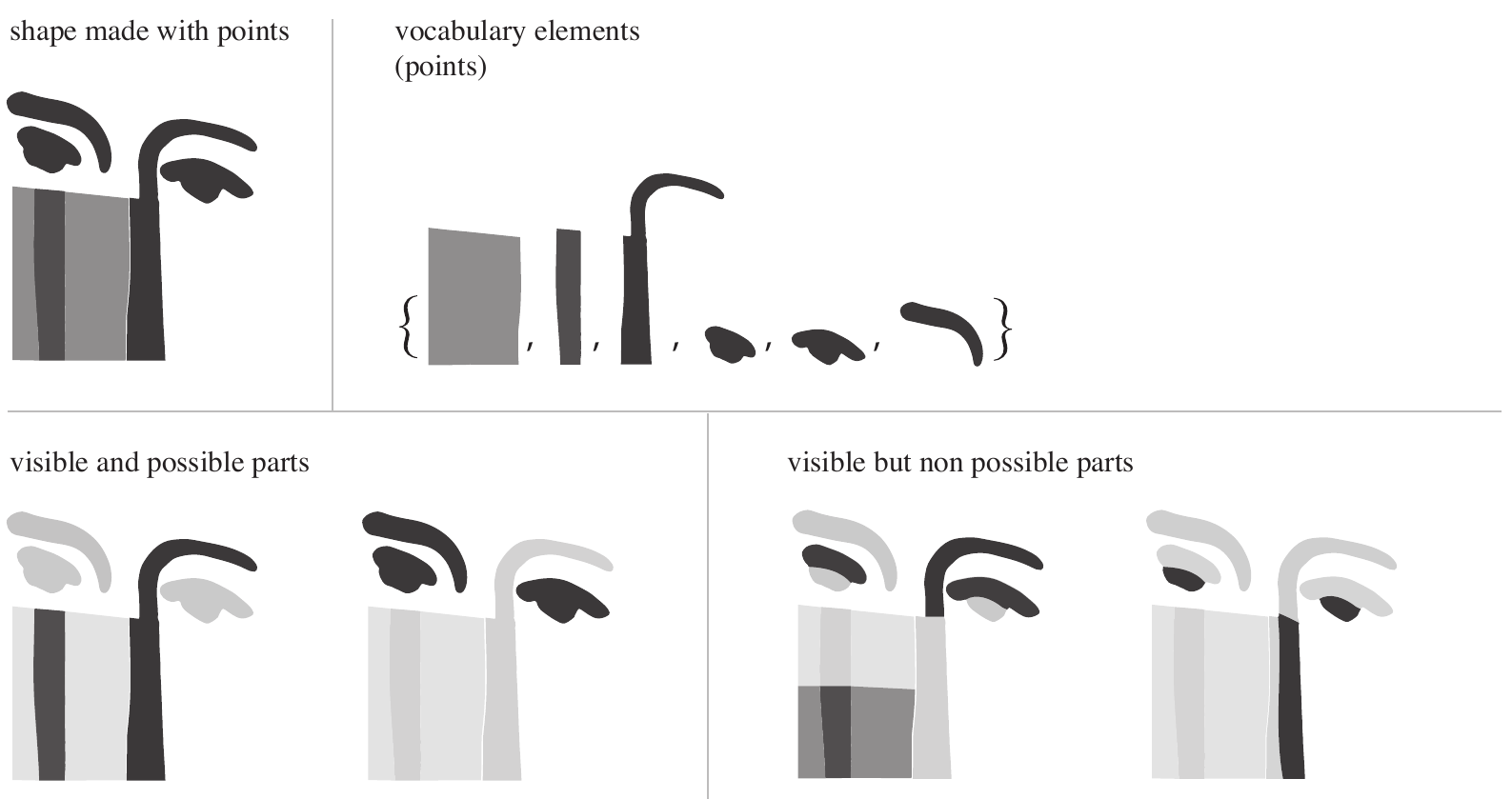}
\caption{Shape made with points (top/left), and the vocabulary elements it is composed of (top/right). The possible parts of the shape are those that can be formed by combining one or more points of the shape (bottom/left). There are visible parts of the shape that are not possible to be formed by combinations of points (bottom/right).}
\end{figure}

Returning now to the main topic of this paper, the fact that shapes made with points can be interpreted into parts only in terms of subsets of points, makes topology on shapes made with points no different than topology on finite sets. In particular, a topology on a shape made with points is solely a matter of inducing an appropriate structure on the set of points out of which the shape is made of. In this sense, the study of topological structure naturally coincides with the study of \emph{preorder relations} on points. After establishing this fact in the next section, some connections between topologies for shapes made with points ($U_0$) and topologies for point-free shapes ($U_i$, for $i > 0$) are discussed, and the main differences between the two are summarized.

In the remaining paragraphs, $S$ represents a shape made with $n$ number of points, for some natural number $n > 0$, and $\mathcal{T}$ a nonempty (finite) set of parts of $S$, satisfying the three basic requirements for a topology given in \cite{R3}. The operations of sum (+) and product ($\cdot$) that make up the algebraic structure of $\mathcal{T}$, coincide with union and intersection of sets. Moreover, the part relation for shapes ($\leq$) coincides with the subset ($\subset$) relation for sets and the statement that $p \leq S$, for any point $p$ of $S$, has essentially the same meaning as the set-theoretic containment relation, that is, $p \in S$.

\section{Order topology on a shape made with points}
A finite topology and a preorder on $S$ are mutually inverse constructions. Start with a preorder relation on the points of $S$ and derive a topology $\mathcal{T}$ out of this relation or, conversely, start with a topology $\mathcal{T}$ for $S$ (by presenting the open parts) and derive a preorder relation on the points of $S$ based on this topology. In other words, a finite topology and a preorder on $S$ represent the same combinatorial object, considered from two different perspectives. A technical explanation of this argument now follows.

Suppose $S$ is equipped with a preorder (reflexive and transitive relation), denoted with the symbol $\preceq$. For each point $p$ of $S$, the \emph{minimal open part} that contains it is given by,


\begin{align}
U_p = \sum q, \; \textrm{for all}\;q \preceq p.
\end{align} 

\noindent (By reflexivity, $p$ is in $U_p$). Let $\mathcal{U}$ = $\{U_p\}_{p \in S}$ be the collection of all minimal open parts containing each of the points of $S$. Then $\mathcal{U}$, augmented with the \emph{empty shape} (\emph{0}), constitutes a \emph{basis} that generates a finite topology $\mathcal{T}$ on $S$. We call this topology the \emph{order topology} on $S$. The collection $\mathcal{U}$ is a unique minimal basis for $\mathcal{T}$.

In the opposite direction, suppose $S$ is equipped with a topology $\mathcal{T}$. For each point $p$ of $S$, the minimal open part $U_p$ that contains it is equal to the \emph{product} of all open parts in $\mathcal{T}$ that contain $p$. 

Now, let $p$ and $q$ be any two points of $S$ and suppose $U_p$ and $U_q$ are the minimal open parts that contain them. Define a preorder $\preceq$ on $S$ by,

\begin{align}
q \preceq p \;\;\textrm{if}\; q \in U_p.
\end{align}

\noindent A working example, demonstrating the use of (1) and (2), is given at the end of this paper.

As discussed in \cite{R3}, many classical concepts for topological spaces will not always apply to shape topologies in a natural manner. Many concepts may not be relevant in general, especially those expressed in terms of points. This is not, however, the case with topology for shapes made with points. Classical topological concepts, including separability, connectedness, continuity, homotopy type and others that have been worked out for finite topological spaces can be adapted for shapes made with points from $U_0$, without significant alterations. The relevant topological toolkit can be found in the canonical literature on finite algebraic and combinatorial topology (\emph{e.g.} \cite{R1}; \cite{R13}; \cite{R10}; \cite{R8}; \cite{R2}). A few basic examples follow for purposes of demonstration only; these are not much different from the equivalent constructions for finite spaces.

An order topology $\mathcal{T}$ for $S$ can be said to satisfy the T$_0$ separation axiom if, and only if, the preorder relation implied by $\mathcal{T}$ is a partial order (that is to say, if the preorder is also antisymmetric). To decide if a given order topology $\mathcal{T}$ for $S$ is indeed T$_0$, one may use the following well-known criterion (adapted from \cite{R10}): $\mathcal{T}$ is T$_0$ if for any two points $p$ and $q$ of $S$, $U_p$ = $U_q$ implies $p$ = $q$ (see also the \emph{Example} for an illustration.) An order topology $\mathcal{T}$ is said to be \emph{discrete} if, and only if, all points of $S$ are open and closed at the same time in $\mathcal{T}$, in which case $\mathcal{T}$ can also be said to satisfy the T$_1$ separation axiom. Note the notion of discreteness here has the same meaning as in topologies for sets. To decide if an order topology is discrete, one may use the following criterion: $\mathcal{T}$ is discrete if, and only if, for all points $p$ of $S$ it holds that $p$ = $U_p$ (the point itself is the smallest open part containing it).

When the points of a shape are known and given in advance, to define a topology one need only \enquote{interrelate} these points in a certain special way. For design applications, the following two ways of defining topologies for shapes made with points appear relevant.

Using a set grammar (this special concept of grammar I am referring to is described in \cite{R11}), define as points the distinct compositional parts (with or without labels) that participate in spatial relations given in the rules of the grammar. A design generated by a set grammar is essentially a shape made with points. Induce an order topology on a design in the following manner: define a preorder relation over the points of the design based on the sequence of rule applications followed to generate it (\emph{e.g.} say that $p \preceq q$ if the point $q$ is generated by a rule applied to $p$.); then, the topology follows from the construction described in (1).

Another way is to use graph representations of shapes. After representing a shape as a transitively oriented directed graph (\cite{R5}), an order topology can be generated on the shape based on this graph, by establishing a preorder relation as follows: $q \preceq p$, if there is a directed edge ($p, q$) in the graph of the shape, going from vertex $p$ to vertex $q$ (i.e. if $q$ is adjacent to $p$). Thereafter, the minimal open parts for all points of $S$ can be calculated using (1). Similar methods of constructing order topologies based on graphs are described, for example, in \cite{R9} for modeling the structure of protein molecules, and in \cite{R7} for the design of low-level image processing algorithms.

What happens with order topology when one does not work with points to begin with? Put differently, is it possible to somehow define an order topology on unanalyzed shapes (when $i > 0$)? A straightforward way to do this is to convert a pre-existing topology for the shape into an order topology. This conversion amounts to constructing a certain kind of mapping (a \emph{many-to-one} transformation), which is sketched in the following paragraph.

Suppose that $C$ is a shape in an algebra $U_i$, for $i > 0$, and is induced with a shape topology $\mathcal{T}_C$. Further, let $\mathcal{B}$ = \{$b_1$,..., $b_k$\} be the reduced (minimal) basis that generates $\mathcal{T}_C$, for some natural number $k$. Then, shape $C$ can be cast into a set of points relative to $\mathcal{T}_C$ by considering as \enquote{points} the basis elements of $\mathcal{B}$ (\cite{R3}).

Let $\mathcal{M}$ be the relation matrix corresponding to $\mathcal{T}_C$. A preorder relation $\preceq$ can be defined on the “points” of $C$ based on $\mathcal{M}$ by,

\begin{align}
b_i \preceq b_j \;\;\textrm{if}\; m_{ij} = 1
\end{align} 

\noindent where $m_{ij}$ is the value of matrix $\mathcal{M}$ at index ($i, j$). Using the preorder obtained by (3), define the minimal open parts containing each of the points of $C$ using (1), and from there, generate an order topology $\mathcal{T}$ for $C$ in a manner already described.

This procedure maps any shape topology $\mathcal{T}_C$, with $k$ number of basis elements, into a corresponding order topology $\mathcal{T}$, with $k$ number of points. It is not so hard to see that the order topology obtained from this procedure must be at least T$_0$. Every relation matrix for a shape topology corresponds to a partial order (immediate from the fact that finite topologies for shapes when $i > 0$ are essentially partially-ordered sets). Thus, the preorder $\preceq$ obtained by (3) is also a partial order by construction, so that $\mathcal{T}$ must be a T$_0$ topology. $\mathcal{T}$ may additionally satisfy the T$_1$ separation axiom if, and only if, the original shape topology $\mathcal{T}_C$ is totally disconnected.

Two differences between shape topologies for unanalyzed shapes and order topologies for shapes made with points are immediate. For a shape made with $n$ number of points, there is both a smallest and a largest finite topology\textemdash the indiscrete topology, with only two open parts, and the discrete topology with 2$^n$ open parts. For an unanalyzed shape, there is a smallest finite topology (namely, the indiscrete one) but there is no largest finite topology. Moreover, a shape made with points has a fixed and absolute underlying space, namely, the set of its points. This space will not change under different topologies\textemdash a new topology will only change how points are structured into subsets, but never the points themselves. On the other hand, there is no fixed underlying space for a shape without points; such a space can be defined only relative to a shape topology and it changes under different shape topologies.

There is another noticeable gulf between doing topology on shapes without points ($i > 0$) and doing topology on shapes made with points ($i = 0$). The former are determined in an open ended interpretative process, where one has to choose which parts of a shape to recognize in order to define a topological structure\textemdash there are indefinitely many ways to do this, since the appearance of a shape without points can be interpreted in indefinitely many \emph{different} ways. The latter are determined by different ways of structuring the same finite set of points, which is given in advance\textemdash there are only finitely many ways to do this, since there can be finitely many different topologies on the same set of points (e.g. see \cite{R4}).

In conclusion, this paper along with the accompanying study in \cite{R3} should give a unified portraiture of finite topology for shapes formed with basic elements of the algebras $U_i$, for $i \geq 0$. When all that matters in a particular investigation is how a set of fixed and indivisible points of a shape are interrelated with one another, then order topology becomes a usual pick. On the other hand, if an investigation is concerned with the more open-ended interface between the structure and appearance of shapes, then shape topologies are the more appropriate choice.\vspace{0.2in}

\begin{figure}[b]
\includegraphics[width=\textwidth]{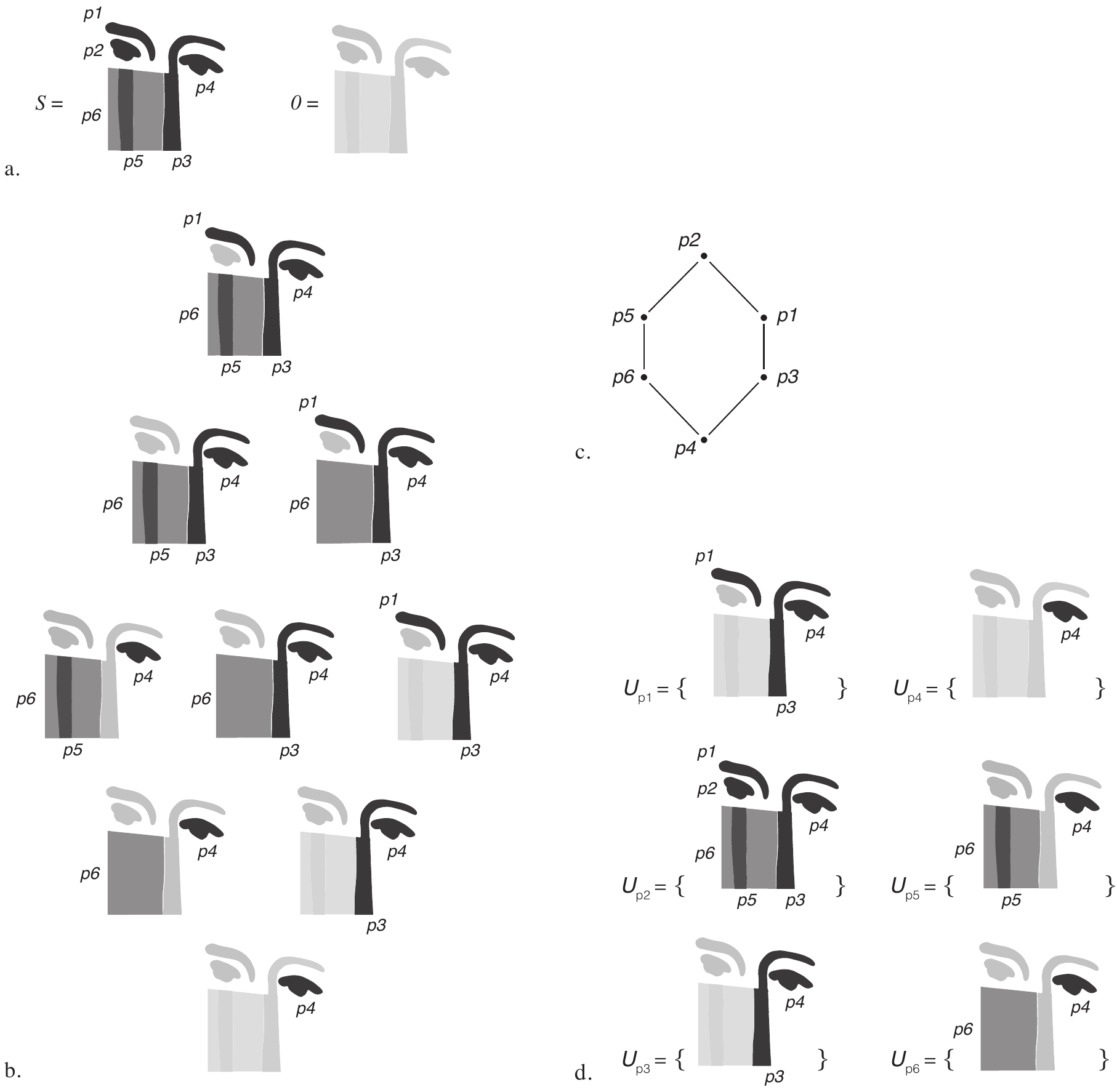}
\caption{(a) Shape made with points (\emph{S}) and \emph{empty shape} (\emph{0}). (b) An order topology $\mathcal{T}$ on the shape. (c) A lattice diagram describing the order relations between the points of the shape based on $\mathcal{T}$. (d) Minimal open parts for each of the points of the shape.}
\end{figure}

\noindent \emph{EXAMPLE \;\;\; Let $S$ = \{$p1$, $p2$,..., $p6$\} be the shape in Figure 5a formed with the points given in Figure 4; the labelling of the points of $S$ is arbitrary. (Note a labelled and a non-labelled shape made with points have no actual mathematical differences. The labelling here is only for convenience, so that it is easier to reference the points of the shape.) To specify a topology on $S$, build a collection $\mathcal{T}$ of recognized parts of $S$, combining points of $S$, such that the resulting collection satisfies the conditions for a topology\textemdash this is shown in Figure 5b. In addition to the parts shown in Figure 5b, $\mathcal{T}$ must also contain $S$ itself and the empty shape (shown in Figure 5a). The minimal open parts containing each point pi of $S$ are shown in Figure 5d; for each point, the minimal open part is the product of open parts in $\mathcal{T}$ that contain it. Thereafter, the preorder relation implied on $S$ based on $\mathcal{T}$ can be defined using the statement in (2): p4 $\preceq$ p6, p4 $\preceq$ p3, p6 $\preceq$ p5, p3 $\preceq$ p1, p1 $\preceq$ p2, and p5 $\preceq$ p2. These are summarized with the lattice diagram shown in Figure 5c. If (pi, pj) is an edge of the diagram, then pi $\preceq$ pj (this means pj \enquote{covers} pi) and $U_{pi} \subset U_{pj}$. One may also want to confirm the minimal open parts shown in Figure 5d using the statement in (1). Topology $\mathcal{T}$ satisfies the T$_0$ separation axiom: $U_{pi}$ = $U_{pj}$ implies pi = pj, for any two points pi and pj of $S$. Thus, the preorder obtained from $\mathcal{T}$ on the points of $S$ is a partial order.}

\begin{dci}The Author declares that there is no conflict of interests.
\end{dci}

\begin{funding}
This research received no specific grant from any funding agency in the public, commercial, or not-for-profit sectors.
\end{funding}

\begin{acks}
This is a preprint of an article published in \emph{Environment and Planning B: Urban Analytics and City Science} (SAGE), available at: \url{https://doi.org/10.1177/2399808319827015}{}.
\end{acks}


\end{document}